\input{amssymb.sty}
\documentstyle[aps,prl,floats,epsf,epsfig]{revtex}

\def\Tc{T_{\mbox{\scriptsize c}}}
\def\rhoc{\rho_{\mbox{\scriptsize c}}}
\begin{document}

\twocolumn[\hsize\textwidth\columnwidth\hsize\csname@twocolumnfalse\endcsname 

\title{Discretization Dependence of Criticality in Model Fluids: a Hard-core Electrolyte}

\author{Young C.\ Kim and Michael E.\ Fisher}
\address{Institute for Physical Science and Technology,\\ University of Maryland, College Park, Maryland 20742}

\date{\today}

\maketitle
\begin{abstract}
Grand canonical simulations at various levels, $\zeta=5$-$20$, of fine-lattice discretization are reported for the near-critical 1:1 hard-core electrolyte or RPM. With the aid of finite-size scaling analyses it is shown convincingly that, contrary to recent suggestions, the universal critical behavior is independent of $\zeta$ $(\gtrsim 4)$; thus the continuum $(\zeta\rightarrow\infty)$ RPM exhibits Ising-type (as against classical, SAW, XY, etc.) criticality. A general consideration of lattice discretization provides effective extrapolation of the {\em intrinsically} erratic $\zeta$-dependence, yielding $(\Tc^{\ast},\rhoc^{\ast})\simeq (0.0493_{3},0.075)$ for the $\zeta=\infty$ RPM.\\
\\
$~~~~~~~~~~~~~~~~~~~~~~~~~~~~~~~~~~~~~~~~~~~~~~~~~~~~~~$ PACS~ numbers:~ 64.60.Fr,~ 02.70.Rr,~ 05.70.Jk,~ 64.70.Fx
\end{abstract}
\vspace{0.2in}
]

The nature of Coulombic criticality has been an outstanding experimental and theoretical issue for more than a decade \cite{fis,wei:sch}. Early experiments suggested that some electrolytes exhibit classical or van der Waals (vdW) critical behavior while others display Ising-type criticality. These results raise the central question: Is Coulombic criticality of vdW, Ising, or other type? In contrast to simple nonionic fluids, in which experiments, theory, and simulation point unequivocally towards Ising behavior, the situation in electrolytes is more challenging owing to the subtle interplay between strong but screened long-range ionic forces and the diverging critical density fluctuations.

Nevertheless, recent experiments favor Ising-type criticality \cite{wei:sch,wie}, as do simulations \cite{cai:lev:wei,ork:pan,yan:pab,lui:fis:pan,kim:fis:lui} of the simplest 1:1 equisized hard-core electrolyte model --- the so-called restricted primitive model or RPM \cite{fis,wei:sch}. On the other hand, the recent, most precise simulations \cite{lui:fis:pan,kim:fis:lui} were performed on a ``finely discretized'' or lattice version of the model at the relatively low discretization level of $\zeta$$\,=\,$$5$. Moreover, in 1998 Valleau and Torrie \cite{val:tor} claimed: ``So far the results offer no support for the existence of any simple Ising-type behavior'' in the (more realistic) continuum model, while in 2002 they asserted that ``the behaviors of the continuum and discretized models are {\em strikingly dissimilar}.'' The work reported here aims to address this specific issue by studying simulations of the RPM at increasing levels of discretization and, more generally, to elucidate discretization effects and to show how one may effectively extrapolate to the continuum limit $\zeta$$\,\rightarrow\,$$\infty$.

For completeness, we recall, first, some details of the RPM: $N$ equisized hard spheres of diameter $a$ in a volume $V$, half carrying charge $+q_{0}$ and half $-q_{0}$, interact via the Coulomb potential $\pm q_{0}^{2}/Dr$ in a medium (representing solvent) of dielectric constant $D$. Dimensionless reduced density and temperature variables are then
  \begin{equation}
    \rho^{\ast} = Na^{3}/V= \rho a^{3},  \hspace{0.3in} T^{\ast}=k_{\mbox{\scriptsize B}}TDa/q_{0}^{2}.  \label{eq1}
  \end{equation}
The model exhibits phase separation into two neutral phases, ion rich and ion poor, at $\Tc^{\ast}$$\,\simeq\,$$\frac{1}{20}$: see Table I.
\begin{table}
\vspace{0.05in}
 \caption{ MC estimates of $\Tc^{\ast}(\zeta)$ and $\rhoc^{\ast}(\zeta)$ for the RPM.   \label{table1}}
\begin{tabular}{ccc|ccc} 
    Ref. $(\zeta$$=$$\infty)$  &  $10^{2}\Tc^{\ast}$   & $10^{2}\rhoc^{\ast}$ & $\zeta$ & $10^{2}\Tc^{\ast}$ & $10^{2}\rhoc^{\ast}$ \\  \hline
  1996 [4(a)] & $4.87(1)$  & $6.5(2)$ & 5  &  5.069(2) & 7.90(25) \\ 
  1999 [4(b)] & $4.88(2)$  & $8.0(5) $ & 8  & 4.966(2) & 7.60(20) \\ 
  1999 [5(b)] & $4.90(3)$  & $7.0(5)$  & 10 & 4.952(5) & 7.60(20) \\
  1999 [6]    & $4.92(3) $ & $6.2(5)$  & 15 & 4.948(5) & 7.55(20) \\
  2002 [13(c)] & $4.89(3) $ & $7.6(3)$  & 20 & 4.940(5) & 7.50(20) \\ 
  2002 [4(c)] & $4.917(2)$ & $8.0(5)$  & $\infty$ & 4.933(5) & 7.50(10)\\ 
\end{tabular}
\end{table}

 Near criticality, the Debye length $\xi_{D}$$\,=\,$$\sqrt{T^{\ast}a^{2}/4\pi\rho^{\ast}}$ is small, $\sim\,$$\frac{1}{4}a$, and many tightly bound neutral clusters form: these cause problems both for theory and simulation. 

Approximate theories yield classical critical exponents \cite{fis:lev}; conversely, simulations with finite $N$ and $V$ exhibit rounded critical points so that finite-size scaling techniques are needed to extract reliable conclusions. Monte Carlo (MC) simulations through 2001 indicate $\rhoc^{\ast}$$\,=\,$$0.060$-$0.085$: see Table I; however, all these values have been derived by {\em assuming} Ising-type criticality and employing the mixed-field finite-size scaling method \cite{bru:wil}. Although this approach neglects possibly significant pressure-mixing terms [11(b),12], the crucial point here is that even though these simulations exhibit {\em consistency} with Ising-type criticality, they do {\em not} rule out other candidates, such as vdW, XY, etc. \cite{lui:fis:pan}.

On the other hand, in recent work, {\bf LFP} [7(b)] convincingly resolved Ising-type critical behavior from other `nearby' candidates via extensive grand canonical simulations; but, for computational efficiency, {\bf LFP} studied only the finely discretized, $\zeta$$\,=\,$$5$ version of the RPM, the discretization level being defined by $\zeta$$\,\equiv\,$$a/a_{0}$ with $a_{0}$ the spacing of the `imposed' lattice \cite{pan:kum}. Indeed, the speed of such lattice simulations can be 100 times faster than off-lattice `continuum' calculations \cite{pan:kum}. Now, when $\zeta$$\,\rightarrow\,$$\infty$, the discretized model clearly approaches the standard continuum RPM; conversely, for $\zeta$$\,=\,$$1$, it corresponds to the most basic lattice model that excludes only double occupancy of individual sites. Nevertheless, for $\zeta$$\,\geq\,$$3$, one finds that the discretized models have phase diagrams with gas-liquid separation that are qualitatively quite similar to the continuum RPM \cite{pan:kum}. 

To conclude that the $\zeta$$\,=\,$$5$ RPM belongs to the $(d$$\,=\,$$3)$-dimensional Ising (or $n$$\,=\,$$1$) universality class, {\bf LFP} first estimated the critical point precisely using multihistogram reweighting and unbiased finite-size scaling methods \cite{ork:fis:pan,kim:fis}. Then, via unbiased extrapolation techniques \cite{ork:fis:pan,kim:fis}, they estimated the critical exponents $\gamma$ and $\nu$ obtaining $1.24(3)$ and $0.63(3)$. These agree well with the Ising values, $\gamma$$\,=\,$$1.23_{9}$ and $\nu$$\,=\,$$0.630_{3}$, and exclude not only vdW criticality (with $\gamma$$\,=\,$$1$, $\nu$$\,=\,$$\frac{1}{2}$), but also XY ($n$$\,=\,$$2$: with $\gamma$$\,\simeq\,$$1.316$, $\nu$$\,\simeq\,$$0.670$), self-avoiding walk (SAW or $n$$\,=\,$$0$: with $\gamma$$\,\simeq\,$$1.159$, $\nu$$\,\simeq\,0.588$), and $n$$\,=\,$$1$ criticality with long-range potentials $|\varphi(r)|$$\,<\,$$\Phi/r^{4.9}$. 

Now, if one accepts, say on renormalization-group theoretical grounds, that universal critical behavior is independent of detailed features of a system (as {\bf LFP} tacitly assumed), Ising universality may be considered as established for the continuum RPM. But, in light of \cite{val:tor} specifically, or more generally, how much faith may be put on this presupposition --- as yet untested in a complex fluid like the RPM? That is the question we answer here, as well, incidentally, as obtaining improved estimates of $\Tc$ and $\rhoc$ for the continuum RPM: see Table I.

In particular, to support Ising-type criticality in the continuum RPM, we have studied the model at discretization levels $\zeta$$\,=\,$$5$, $8$, $10$, $15$ and $20$ via grand canonical MC simulations in cubic boxes of dimensions $L^{d}$ with periodic boundary conditions, and sizes varying from $L^{\ast}$$\,\equiv\,$$L/a$$\,=\,$$5$ to $12$. As seen in {\bf LFP}, obtaining several independent confirmations of the class of criticality for a nontrivial system requires a large computational effort. On the other hand, the universality class can be determined with confidence by evaluating sufficiently precisely one universal parameter, say either a critical exponent or an amplitude ratio, that distinguishes readily among reasonable candidates. 

Following {\bf LFP} \cite{lui:fis:pan,kim:fis}, and previous applications to simpler, symmetric systems \cite{bin}, we thus focus on the grand canonical finite-size parameter $Q_{L}$ defined by the dimensionless moment-ratio
  \begin{equation}
   Q_{L}(T;\langle\rho\rangle_{L}) \equiv \langle m^{2}\rangle_{L}^{2}/\langle m^{4}\rangle_{L}, \hspace{0.15in} m = \rho - \langle\rho\rangle_{L},   \label{eq2}
  \end{equation}
where $\langle\cdot\rangle_{L}$ denotes a grand canonical expectation value at fixed $T$ and chemical potential $\mu$ adjusted to yield the mean density $\langle\rho\rangle_{L}$. As well known, $Q_{L}$$\,\rightarrow\,$$\frac{1}{3}$ when $L$$\,\rightarrow\,$$\infty$ in any single-phase region, while $Q_{L}$$\,\rightarrow\,$$1$ on the coexistence curve diameter, $\rho_{\mbox{\scriptsize diam}}(T)$$\,\equiv\,$$\frac{1}{2}(\rho^{+}+\rho^{-})$, where $\rho^{\pm}(T)$ denotes the densities of the coexisting liquid and gas phases. 

But the crucial point here is that $Q_{L}(\Tc,\rhoc)$ approaches a universal value, $Q_{\mbox{\scriptsize c}}$, that serves to resolve distinct criticality classes rather sharply. Thus for Ising systems one has $Q_{\mbox{\scriptsize c}}^{\mbox{\scriptsize Is}}$$\,=\,$$0.623_{6}$ and, as discussed in {\bf LFP}, the classical, SAW and XY values are $Q_{\mbox{\scriptsize c}}^{\mbox{\scriptsize vdW}}$$\,=\,$$0.4569\cdots$, $Q_{\mbox{\scriptsize c}}^{\mbox{\scriptsize SAW}}$$\,=\,$$0$, and $Q_{\mbox{\scriptsize c}}^{\mbox{\scriptsize XY}}$$\,=\,$$0.804_{5}$, while for long-range, $1/r^{3+\sigma}$ Ising systems, $Q_{\mbox{\scriptsize c}}(\sigma)$ increases almost linearly from vdW to Ising values in the interval $\frac{3}{2}$$\,\leq\,$$\sigma$$\,\leq\,$$1.96_{6}$ with $Q_{\mbox{\scriptsize c}}(\sigma$$=$$1.9)$$\,\simeq\,$$0.600$.

To progress it is necessary to calculate $Q_{L}$ along the $Q$-loci, $\rho_{Q}(T;L)$, defined by the value of $\langle\rho\rangle_{L}$ for which $Q_{L}$ is maximal at fixed $T$ [7(b),15]: high precision is essential \cite{kim}. Then finite-size scaling theory \cite{kim:fis} implies that as $L$ increases, successive self-intersection points, say $\Tc^{Q}(L)$, approach the critical point, $\Tc$, rapidly as $1/L^{(1+\theta)/\nu}$: see Fig.\ 1.
\begin{figure}[h]
\vspace{-0.9in}
\centerline{\epsfig{figure=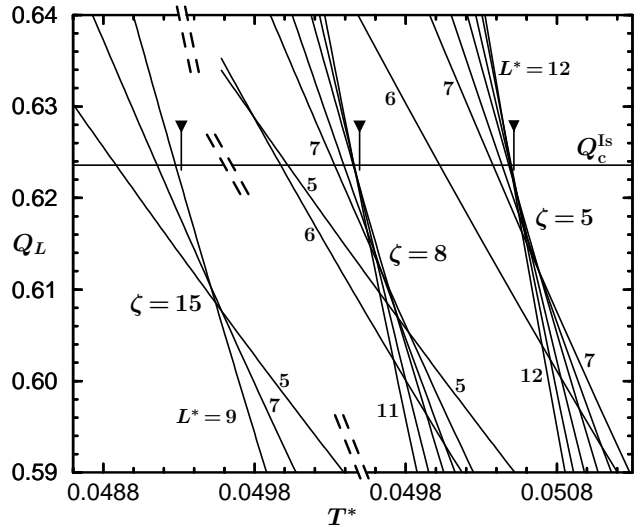,width=3.8in,angle=0}}
\vspace{-1.0in}
\caption{Plots of $Q_{L}$ on the $Q$-loci, $\rho_{Q}(T;L)$, vs.\ $T^{\ast}$ for $\zeta$$\,=\,$$5,8$ and $15$. The horizontal line marks $Q_{L}$$\,=\,$$Q_{\mbox{\scriptsize c}}^{\mbox{\scriptsize Is}}$.  \label{fig1}}
\end{figure}
 In addition, the difference $Q_{L}\mbox{\boldmath $($}T_{\mbox{\scriptsize c}}^{Q}(L);\rho_{Q}\mbox{\boldmath $)$}-Q_{\mbox{\scriptsize c}}$ varies, in leading orders, as $L^{-\theta/\nu}$ followed by a $j_{2}^{2}L^{-2\beta/\nu}$ term \cite{kim:fis}, where $j_{2}$ is the pressure-mixing coefficient \cite{kim:fis:ork}.

As {\bf LFP} observed, the self-intersection points $\Tc^{Q}(L)$ for the $\zeta$$\,=\,$$5$ RPM almost coincide with the universal Ising value providing {\em both} confirmation of Ising character {\em and} a precise estimate of $\Tc$: see Fig.\ 1. Except for a translation by $\Delta T^{\ast}$$\,\simeq\,$$0.0010$, the corresponding plots shown in Fig.\ 1 for $\zeta$$\,=\,$$8$ are almost identical and provide the same results. For $\zeta$$\,=\,$$15$ (as for $\zeta$$\,=\,$$10$ and $20$, not shown) the trends are very similar. Since the values $Q_{\mbox{\scriptsize c}}^{\mbox{\scriptsize vdW}}$ and $Q_{\mbox{\scriptsize c}}^{\mbox{\scriptsize XY}}$ are off scale and $Q_{\mbox{\scriptsize c}}$$\,\leq\,$$0.600$ is implausible, vdW, XY, and long-range Ising criticality with $\sigma$$\,\leq\,$$1.9$ are again excluded for all these values of $\zeta$. Furthermore, since the behavior as $\zeta$ increases by a factor of $4$ changes so little, there seem no grounds to doubt that Ising behavior prevails for all $\zeta$$\,\rightarrow\,$$\infty$.

As a consistency check, consider the slopes, say $Q_{\mbox{\scriptsize c}}^{\prime}(L)$, of the plots at the points where $Q_{L}$$\,=\,$$Q_{\mbox{\scriptsize c}}^{\mbox{\scriptsize Is}}$: by finite-size scaling, these should diverge as $L^{1/\nu}$. Accordingly, in Fig.\ 2(a)
\begin{figure}[h]
\vspace{-0.9in}
\centerline{\epsfig{figure=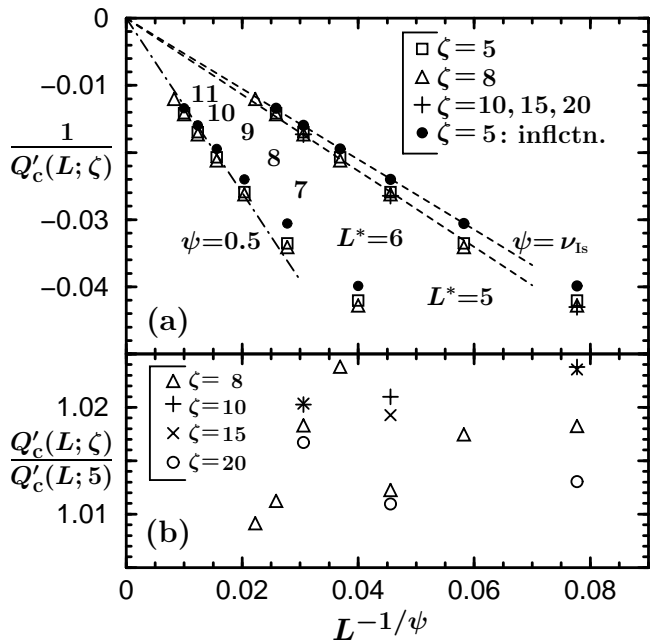,width=4.0in,angle=0}}
\vspace{-0.8in}
\caption{(a) Inverse of the slope, $1/Q_{\mbox{\scriptsize c}}^{\prime}$, evaluated at $Q_{L}$$\,=\,$$Q_{\mbox{\scriptsize c}}^{\mbox{\scriptsize Is}}$ vs.\ $L^{-1/\psi}$ with $\psi$$\,=\,$$\nu_{\mbox{\scriptsize Is}}$ and $\nu_{\mbox{\scriptsize vdW}}$. The solid circles are the inverse of the slopes at the inflection points for $\zeta=5$. (b) Ratios, $Q_{\mbox{\scriptsize c}}^{\prime}(L;\zeta)/Q_{\mbox{\scriptsize c}}^{\prime}(L;\zeta=5)$, vs.\ $L^{-1/\nu_{\mbox{\tiny Is}}}$ for $\zeta=8,10,15$ and 20. \label{fig2}}
\end{figure}
 the inverses $1/Q_{\mbox{\scriptsize c}}^{\prime}(L)$ for $\zeta$$\,=\,$$5$, $8$, $10$, $15$, and $20$ are plotted versus $L^{-1/\psi}$ using $\psi$$\,=\,$$\nu_{\mbox{\scriptsize Is}}$$\,\simeq\,$$0.63$ and $\nu_{\mbox{\scriptsize vdW}}$$\,=\,$$0.5$, along with $1/Q_{0}^{\prime}(L)$, the inverse slope at the {\em inflection points} for $\zeta$$\,=\,$$5$. Evidently, setting $\psi$$\,=\,$$\nu_{\mbox{\scriptsize vdW}}$ does {\em not} satisfactorily capture the asymptotic behavior of the slopes. On the other hand, the slopes at the inflection points for $\zeta$$\,=\,$$5$ (solid circles) provide definite evidence for Ising character while the slopes, $Q_{\mbox{\scriptsize c}}^{\prime}$, for $\zeta$$\,\geq\,$$5$ also support Ising criticality. Perhaps the most striking fact is that the slopes are so insensitive to $\zeta$: indeed, as seen in Fig.\ 2(b), the values of $Q_{\mbox{\scriptsize c}}^{\prime}(L;\zeta)$ for $\zeta$$\,\geq\,$$8$ are no more than $1$ or $2\%$ higher than for $\zeta$$\,=\,$$5$. This striking independence and the clear verdict of Ising criticality for the $\zeta$$\,=\,$$5$ RPM \cite{lui:fis:pan,kim:fis:lui,kim:fis}, reinforces the conclusion that Ising criticality remains valid in the continuum limit.

From the plots of Fig.\ 1, the critical temperature for $\zeta$$\,=\,$$8$ can be estimated with the same precision as for $\zeta$$\,=\,$$5$. Even though only three system sizes have been computed for $\zeta$$\,=\,$$10$, $15$, and $20$, the similar behavior seen in Figs.\ 1 and 2 for different $\zeta$ leads to comparable estimates although with larger uncertainties: see Table I. As found in [13(c)], $\Tc^{\ast}(\zeta)$ falls when $\zeta$ increases through integral values. The critical densities, $\rhoc^{\ast}$, can be estimated by suitably extrapolating the densities $\rho_{Q}(\Tc;L)$, on the $Q$-loci at the (estimated) values of $\Tc^{\ast}$ [7(b),15]. From Table I we see that $\rhoc^{\ast}(\zeta)$ also decreases (in accord with [13(c)]).

Now, to extrapolate effectively to the continuum limit, we ask how $\Tc^{\ast}(\zeta)$, $\rhoc^{\ast}(\zeta)$, etc., should vary when $\zeta$ increases \cite{sar}. To gain insight consider a $d$-dimensional fluid with pair potential $\varphi(\mbox{\boldmath $r$})$$\,=\,$$\varphi_{0}(r)$$\,+\,$$\varphi_{1}(r)$, where $\varphi_{0}$ is repulsive and of short range, say $a$, while $\varphi_{1}$ is smooth, attractive and long ranged. If $B_{i}(T)$$\,=\,$$-\frac{1}{2}\int d^{d}r f_{i}(\mbox{\boldmath $r$})$ with $1+f_{i}$$\,=\,$$e^{-\varphi_{i}(\mbox{\scriptsize\boldmath $r$})/k_{\mbox{\tiny B}}T}$, $i$$\,=\,$$0$, $1$, are partial second virial coefficients, an approximate, vdW-type equation of state is
 \begin{equation}
  p/\rho k_{\mbox{\scriptsize B}}T = [1-\rho B_{0}(T)]^{-1} + \rho B_{1}(T).  \label{eq3}
 \end{equation}
On discretization with $\zeta$$\,=\,$$a/a_{0}$, the $B_{i}$ integrals are replaced by sums, $B_{i}^{\zeta}$$\,\equiv\,$$-\frac{1}{2}\sum_{\mbox{\scriptsize\boldmath $n$}} f_{i}(\mbox{\boldmath $n$} a_{0})$, over integral lattice vectors, $\mbox{\boldmath $n$}$. One must then ask: How rapidly does $B_{i}^{\zeta}$ converge to $B_{i}^{\infty}$? Certainly, the decay of the truncation error, $E_{i}(\zeta)$$\,=\,$$(B_{i}^{\infty}/B_{i}^{\zeta})-1$, when $\zeta$$\,\rightarrow\,$$\infty$, will be dominated by any discontinuities in $f_{i}(\mbox{\boldmath $r$})$ (or its derivatives).

Specifically, for a hard-core $\varphi_{0}$ (as needed for the RPM) one has $B_{0}^{\infty}/B_{0}^{\zeta}$$\,=\,$$V(\zeta)/N(\zeta)$, where $V(\zeta)$ is the volume of a sphere of radius $\zeta$ while $N(\zeta)$ is the number of lattice sites satisfying $|\mbox{\boldmath $n$}|$$\,\leq\,$$\zeta$. A heuristic argument \cite{sar} indicates that $E_{0}$ should be of rms magnitude $c_{d}/\zeta^{(d+1)/2}$ with $c_{d}$$\,=\,$${\cal O}(1)$. This is exact for $d$$\,=\,$$1$ and $2$ and is supported numerically for $d$$\,=\,$$3$ by the scaled plot in Fig.\ 3(a) \cite{ble:dys}.
\begin{figure}[h]
\vspace{-0.6in}
\centerline{\epsfig{figure=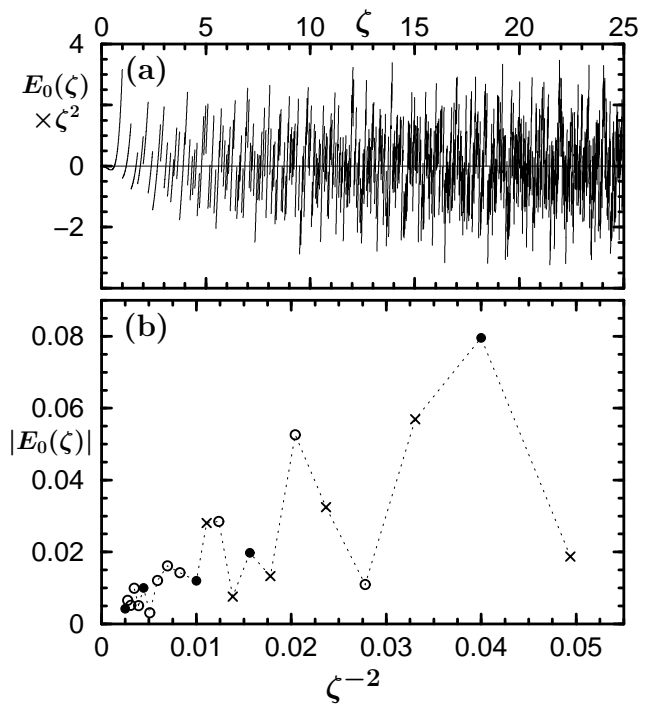,width=4.0in,angle=0}}
\vspace{-0.7in}
\caption{(a) Relative hard-core truncation error, $E_{0}(\zeta)$, for sc lattices: (a) scaled by $\zeta^{2}$; (b) magnitude vs.\ $\zeta^{-2}$ for half-odd $(\times)$ and integral ($\circ$ and $\bullet$) $\zeta$ values. \label{fig3}}
\end{figure}
 Evidently, $E_{0}(\zeta)$ varies wildly and discontinuously; and the noisiness persists when $\zeta$ is restricted to, e.g., half-integers: see Fig.\ 3(b).

Furthermore, as implied by (3) \cite{sar}, the erratic behavior transfers to $\rhoc^{\ast}(\zeta)$ and $\Tc^{\ast}(\zeta)$ and seriously hampers extrapolation: see the plots {\bf (i)} in Fig.\ 4.
\begin{figure}[h]
\vspace{-0.5in}
\centerline{\epsfig{figure=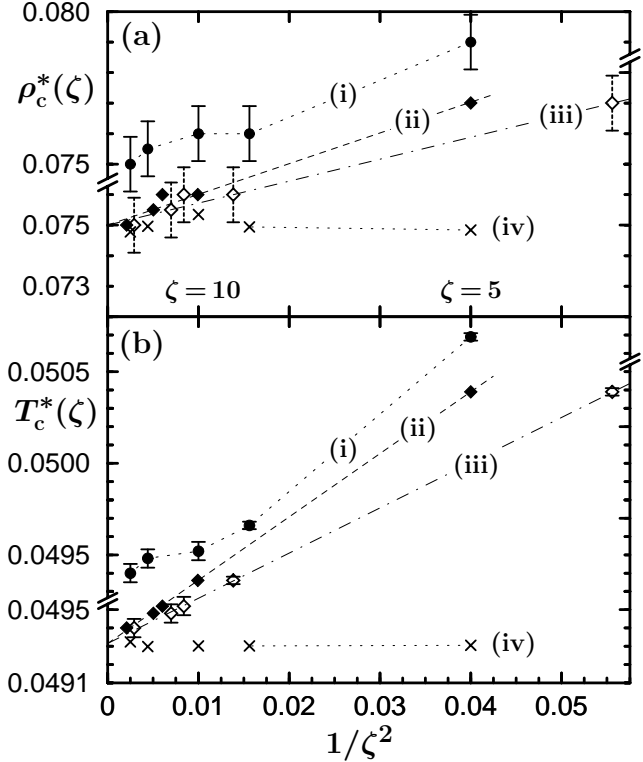,width=4.0in,angle=0}}
\vspace{-0.5in}
\caption{Estimation of $\rhoc^{\ast}$ and $\Tc^{\ast}$ for the RPM. Above scale breaks: {\bf (i)} simulation values vs.\ $\zeta^{-2}$; below breaks: {\bf (ii)} vs.\ $1/\zeta^{\dag\, 2}(\zeta)$; {\bf (iii)} vs.\ $\frac{1}{2}/(\zeta^{\dag}-2)^{2}$; {\bf (iv)} rescaled simulation values, $\rhoc^{\dag}(\zeta)$ and $\Tc^{\dag}(\zeta)$, vs.\ $\zeta^{-2}$: see (4). \label{fig4}}
\end{figure}
 We present two strategies to mitigate the problem. First, define a modified discretization level $\zeta^{\dag}(\zeta)$, via $E_{0}(\zeta)$$\,=\,$$c^{\dag}/\zeta^{\dag\, 2}$: $c^{\dag}$$\,=\,$$25E_{0}(5)$ is convenient. Then, as in Fig.\ 4 plots {\bf (ii)} and, with an $\epsilon$$\,=\,$$2$ shift, {\bf (iii)}, examine $\rhoc^{\ast}(\zeta)$, etc., vs.\ $1/\zeta^{\dag\, 2}$. Evidently, the behavior is much smoother!

Second, an {\em effective} $B_{1}(T)$ for the RPM must include contributions from the Coulombic interactions. Accordingly, in the hope of improving convergence, we supplement $B_{0}^{\zeta}$ by coefficients, $b_{1}$, independent of $\zeta$, and {\em rescale} the discretized densities via
 \begin{equation}
  \rhoc^{\dag}(\zeta) \equiv \rhoc^{\ast}(\zeta)(B_{0}^{\zeta}+b_{1}^{\rho})/(B_{0}^{\infty}+b_{1}^{\rho}),  \label{eq4}
 \end{equation}
and similarly for $\Tc$. Indeed, for the choices $b_{1}^{\rho}$$\,=\,$$0.4B_{0}^{\infty}$ and $b_{1}^{T}$$\,=\,$$1.7B_{0}^{\infty}$ both $\rhoc^{\dag}(\zeta)$ and $\Tc^{\dag}(\zeta)$ become almost insensitive to $\zeta$: see plots {\bf (iv)}.

From the enhanced plots in Fig.\ 4 we estimate $\Tc^{\ast}$$\,\simeq\,$$0.04933$ and $\rhoc^{\ast}$$\,\simeq\,$$0.075$ for the RPM; see Table I. Our value for $\Tc^{\ast}$ agrees well with the (less precise) estimate of \cite{yan:pab}, but their estimate of $\rhoc^{\ast}$ is very low. Other recent estimates of $\rhoc^{\ast}$ encompass our value but the $\Tc^{\ast}$ estimates fall lower \cite{kim2}.

In summary, Monte Carlo studies of $Q_{L}(T)$ on the $Q$-loci of the restricted primitive model (RPM) for various discretization levels, $\zeta$$\,=\,$$5$-$20$, have provided convincing evidence for Ising-type, as against XY, or SAW, etc., criticality in the continuum limit. By pinpointing, generally, the primary sources of discretization errors \cite{sar} we have found effective means of estimating precisely the limiting critical parameters from data for $\zeta$$\,\lesssim\,$$10$.

We are indebted to Erik Luijten for the use of his histogram reweighting program and to Pavel Bleher for informative correspondance. We thank Thanos Panagiotopoulos for his interest and encouragement and the National Science Foundation for support (through Grant No.\ CHE 03-01101).


\vspace{-0.2in}
\begin{thebibliography}{99}
\vspace{-0.5in}
  \bibitem{fis} See, e.g., M.\ E.\ Fisher, J.\ Stat.\ Phys.\ {\bf 75}, 1 (1994); G. Stell, J.\ Stat.\ Phys.\ {\bf 78}, 197 (1995).
  \bibitem{wei:sch} For a recent review, see H.\ Weing\"{a}rtner and W.\ Schr\"{o}er, Adv.\ Chem.\ Phys.\ {\bf 116}, 1 (2001).
  \bibitem{wie} M.\ Kleemeier, {\em et al.} J.\ Chem.\ Phys.\ {\bf 110}, 3085 (1999).
  \bibitem{cai:lev:wei} (a) J.\ M.\ Caillol, D.\ Levesque, and J.\ J.\ Weiss, Phys.\ Rev.\ Lett.\ {\bf 77} 4039 (1996); (b) J.\ Chem.\ Phys.\ {\bf 111}, 9509 (1999); (c) {\bf 116}, 10794 (2002).
  \bibitem{ork:pan} (a) G.\ Orkoulas and A.\ Z.\ Panagiotopoulos, J.\ Chem.\ Phys.\ {\bf 101}, 1452 (1994); (b) {\bf 110}, 1581 (1999).
  \bibitem{yan:pab} Q.\ Yan and J.\ J.\ de Pablo, J.\ Chem.\ Phys.\ {\bf 111}, 9509 (1999).
  \bibitem{lui:fis:pan} E.\ Luijten, M.\ E.\ Fisher and A.\ Z.\ Panagiotopoulos, (a) J.\ Chem.\ Phys.\ {\bf 114}, 5468 (2001); (b) Phys.\ Rev.\ Lett.\ {\bf 88}, 185701 (2002) which we denote as {\bf LFP}.
  \bibitem{kim:fis:lui} Y.\ C.\ Kim, M.\ E.\ Fisher and E.\ Luijten, Phys.\ Rev.\ Lett.\ {\bf 91}, 065701 (2003).
  \bibitem{val:tor} (a) J.\ Valleau and G.\ Torrie, J.\ Chem.\ Phys.\ {\bf 108}, 5169 (1998); (b) {\bf 117}, 3305 (2002).
  \bibitem{fis:lev} See, e.g., Y.\ Levin and M.\ E.\ Fisher, Physica A {\bf 225}, 164 (1996) and references in \cite{fis,wei:sch}.
  \bibitem{bru:wil} (a) A.\ D.\ Bruce and N.\ B.\ Wilding, Phys.\ Rev.\ Lett.\ {\bf 68}, 193 (1992); (b) for a recent assessment, see Y.\ C.\ Kim and M.\ E.\ Fisher, arXiv:cond-mat/0310247.
  \bibitem{kim:fis:ork} Y.\ C.\ Kim, M.\ E.\ Fisher and G.\ Orkoulas, Phys.\ Rev.\ E {\bf 67}, 061506 (2003).
  \bibitem{pan:kum} (a) A.\ Z.\ Panagiotopoulos and S.\ K.\ Kumar, Phys.\ Rev.\ Lett.\ {\bf 83}, 2981 (1999); (b) A.\ Z.\ Panagiotopoulos, J.\ Chem.\ Phys.\ {\bf 112}, 7132 (2000); (c) {\bf 116}, 3007 (2002).
  \bibitem{ork:fis:pan} G.\ Orkoulas, M.\ E.\ Fisher, and A.\ Z.\ Panagiotopoulos, Phys.\ Rev.\ E {\bf 63}, 051507 (2001).
  \bibitem{kim:fis} Y.\ C.\ Kim and M.\ E.\ Fisher, Phys.\ Rev.\ E {\bf 68}, 041506 (2003).
  \bibitem{bin} K.\ Binder, Phys.\ Rev.\ Lett.\ {\bf 47}, 693 (1981); K.\ Binder and D.\ P.\ Landau, Phys.\ Rev.\ B {\bf 30}, 1477 (1984).
  \bibitem{kim} To achieve precise, reliable data we used, for each $L^{\ast}$, a total of 30-55 state points {\bf (}SPs, i.e., $(T^{\ast},\rho^{\ast})$ values{\bf )} in $(0.0487$$\,<\,$$T^{\ast}$$\,<\,$$0.0513$, $0.08$$\,<\,$$\rho^{\ast}$$\,<\,$$0.17)$ with close spacing near criticality. A typical SP had $(2$-$10)$$\times$$10^{4}$ {\em independent} samples amounting to $\sim\,$$10^{10}$ MC steps per SP.
  \bibitem{sar} Our arguments will be expounded more fully and illustrated with Sarvin Moghaddam: see also Chap.\ 7 of her Ph.D. Thesis, University of Maryland, 2003.
  \bibitem{ble:dys} But see P.\ M.\ Bleher and F.\ J.\ Dyson, Acta Arith.\ {\bf 68}, 383 (1994), {\bf 73}, 199 (1995), who show that $c_{3}^{2}={\cal O}(\ln\zeta)$. Furthermore, for all $d>3$ one has $E_{0}\sim 1/\zeta^{2}$.
  \bibitem{kim2} The uncertainty quoted for $\Tc^{\ast}$ in [4(c)] seems unduly small considering the rather noisy data presented; and the number of SPs used is far less than we have employed.
\end{thebibliography}
\end{document}